\documentstyle[12pt]{article}
\newcommand{\rf}[1]{(\ref{#1})}
\newcommand{\beq}{\begin{equation}}
\newcommand{\eeq}{\end{equation}}

\newcommand{\G}{\Gamma}

\renewcommand{\l}{\lambda}
\newcommand{\k}{\kappa}

\renewcommand{\S}{\Omega}
\newcommand{\bea}{\begin{eqnarray}}
\newcommand{\eea}{\end{eqnarray}}

\newcommand{\cO}{{\cal O}}

\newcommand{\cC}{{\cal C}}
\newcommand{\cS}{{\cal S}}
\newcommand{\cM}{{\cal M}}

\newcommand{\cT}{{\cal T}}

\newcommand{\noi}{\noindent}

\renewcommand{\i}{^{-1}} 

\newcommand{\ver}{\thinspace\vert\thinspace}
\newcommand{\la}{\langle}
\newcommand{\ra}{\rangle}
\newcommand{\orbi}{\cC\wr\S}

\begin{document}
\topmargin 0pt
\oddsidemargin 5mm
\headheight 0pt
\topskip 0mm
\addtolength{\baselineskip}{0.20\baselineskip}
\pagestyle{empty}
\hfill 
\begin{center}
\vspace{3 truecm}
{\Large \bf Orbifoldization, covering surfaces \\ 
and uniformization theory}
\vspace{5 truecm}

{\large Peter Bantay}
\vspace{1 truecm}

{\em Institute for Theoretical Physics\\
Rolland E\"otv\"os University, Budapest \\}

\vspace{3 truecm}
\end{center}
\noi
\underbar{\bf Abstract} 
The connection between the theory of permutation orbifolds, covering surfaces
and uniformization is investigated, and the higher genus partition functions
of an arbitrary permutation orbifold are expressed in terms of those of the
original theory.
\vfill
\newpage
\pagestyle{plain}

Consider a Rational Conformal Field Theory $\cC$ and a permutation group
$\S$ of degree $n$. The elements of $\S$ act as global symmetries of the
$n$-fold tensor power $\cC^{\otimes n}$ - whose fields are just $n$-fold
products of fields of $\cC$ -, so one may orbifoldize with respect to $\S$,
obtaining the permutation orbifold $\orbi$. The resulting orbifold theories
have been first considered in \cite{1,2}. More detailed investigations may
be found in \cite{BHS,DM}, the latter giving a mathematically rigorous 
approach through the theory of Vertex Operator Algebras. 
Permutation orbifold techniques have been applied in \cite{DMV2} to the 
computation of the free-energy of the second quantized string, and in
\cite{DV2b} to matrix string theory. The general structure of permutation 
orbifolds for 
arbitrary permutation groups $\S$, including the classification of primary 
fields and the explicit form of the genus 1 characters and their 
modular transformation properties, can be found in \cite{PO1}. An 
important aspect of the results of \cite{PO1} is the connection 
of permutation orbifolds with the theory of covering surfaces.

The aim of this paper is to present a formula expressing the higher genus
partition functions of the permutation orbifold $\orbi$ in terms of the
higher genus partition functions of $\cC$, once again through an application
of the theory of covering surfaces. To achieve this goal, 
let's first take a look at the genus one case.

As explained in \cite{PO1}, the genus one partition function $Z^\S_1(\tau)$ of
$\orbi$ may be expressed through the partition function $Z_1(\tau)$ of $\cC$
as follows :
\beq Z^\S_1(\tau)=\frac{1}{\vert\S\vert}\sum_{{x,y\in \S}\atop{xy=yx}}
\prod_{\xi\in\cO(x,y)} Z_1(\tau_\xi) \label{g1} \eeq
where $\cO(x,y)$ denotes the set of orbits of the group generated by the
commuting permutations $x$ and $y$,   and
\beq \tau_\xi=\frac{\mu_\xi\tau+\k_\xi}{\l_\xi} \eeq
with $\mu_\xi,\k_\xi,\l_\xi$ being numerical characteristics of the orbit $\xi$,
namely $\mu_\xi$ is the number of $x$ orbits contained in $\xi$, $\l_\xi$ is
their common length, while $\k_\xi$ is an integer such that $y^{\mu_\xi}
x^{-\k_\xi}$ belongs to the stabilizer of $\xi$.

Note that the summation in Eq.\rf{g1} over commuting pairs $x,y$ may be
rewritten as a sum over homomorphisms from $\Gamma_1={\bf Z}\oplus {\bf Z}$ 
- the fundamental group of the torus $\Sigma_1$ - into the twist group $\S$,
 $x$ and $y$ corresponding to the images of the generators.
Such a homomorphism $\phi:\G_1\to\S$ determines on one hand a permutation
representation of $\G_1$, and on the other hand an unramified covering
of the surface $\Sigma_1$, where the image under $\phi$ of an element of 
$\G_1$ determines the monodromy around the corresponding curve. This
unramified covering is not connected in general, its connected components
- which are all tori according to Riemann-Hurwitz - being in one-to-one
correspondence with the orbits of the permutation group $\phi(\G_1)$.
Moreover, it may be shown that the modular parameter of the torus
corresponding to some orbit $\xi\in\cO(\phi):=\cO(x,y)$ is just given by
$\tau_\xi$, if the modular parameter of the original torus $\Sigma_1$ 
was $\tau$. In other words
, $Z_1^\S$ is given as a sum over all degree $n$ unramified 
coverings of $\Sigma_1$ whose monodromy group is a subgroup of $\S$,
where each summand is the product  of the contributions of the covering's
connected components.

To attack the higher genus case, our first
task is to discuss the parametrization of Teichmuller space that is the
most appropriate for our purposes. This so-called Fricke parametrization
is connected with the uniformization theory of Riemann surfaces, so we
begin with a sketchy review of this subject ( c.f. \cite{Ima} ).

Consider a closed Riemann surface $\Sigma_g$ of genus $g>1$. 
$\Sigma_g$ is not simply connected, its fundamental group $\G_g$ being 
an infinite group with presentation
\beq \G_g=\la a_1,b_1,\dots ,a_g,b_g\ver \prod_{i=1}^g [a_i,b_i]=1\ra \eeq
where $[a,b]=a\i b\i ab$ denotes the commutator of $a$ and $b$. By the
Riemann mapping theorem, the universal covering of $\Sigma_g$ is the
upper half-plane ${\bf H}=\{z \ver  {\rm Im } z > 0\}$, and 
we have the following isomorphism of Riemann surfaces
\beq \Sigma_g\cong {\bf H}/\tau(\G_g), \eeq
where $\tau : \G_g\to SL(2,{\bf R})$ is an embedding of the fundamental group 
$\G_g$ of $\Sigma_g$ into the group $SL(2,{\bf R}) ={\rm Aut}({\bf H})$ of 
automorphisms of the upper half-plane ${\bf H}$, whose image $\tau(\G_g)$
is a purely hyperbolic Fuchsian group \cite{Bre}. It follows that 
$\tau : \G_g\to SL(2,{\bf R})$ determines a point in the Teichmuller space 
$\cT_g$ of genus $g$ closed surfaces, with the proviso that two embeddings
differing by an inner automorphism of $SL(2,{\bf R})$ correspond to the
same point in $\cT_g$. It should be noted that, if $M_g$ denotes
those automorphisms of the free group generated by
$a_1,b_1,\dots ,a_g,b_g$ that leave the product $\prod_{i=1}^g
[a_i,b_i]$ invariant, then the composite map
\beq \tau\circ\alpha : \G_g \to SL(2,{\bf R}) \eeq
with $\alpha\in M_g$ is an embedding which generally  
corresponds to some other point in
$\cT_g$, but nevertheless induces the same complex structure on the quotient, 
that is $\tau$ and $\tau\circ\alpha$ determine the same point in the
moduli space $\cM_g$ of genus $g$ closed surfaces. This means that
$M_g$ is the genus $g$ mapping class group, with the prescribed action
on $\cT_g$. This coordinatization of $\cT_g$ in terms of the embeddings
$\tau : \G_g\to SL(2,{\bf R})$ is the Fricke parametrization that we shall
use, and a genus $g$ partition function is a function of $\tau\in\cT_g$, 
subject to the requirement of modular invariance
\beq Z(\tau\circ\alpha)=Z(\tau)\qquad\forall \alpha\in M_g \label{mod}\eeq

After these preliminaries, we can express the genus $g$ partition function
$Z_g^\S(\tau)$ of the permutation orbifold $\orbi$ in terms of the
partition functions $Z_g(\tau)$ of $\cC$. The formula reads
\beq 
Z_g^\S(\tau)=\frac{1}{\vert\S\vert}\sum_{\phi : \G_g\to\S}
\prod_{\xi\in\cO(\phi)} Z_{g_\xi}(\tau_\xi) \label{part} \eeq
To understand this formula, note that - in complete analogy with the $g=1$ case
discussed previously - a homomorphism $\phi : \G_g\to\S$
determines a permutation representation of degree $n$ of the group $\G_g$, 
as well as an $n$-sheeted unramified covering of the Riemann surface 
$\Sigma_g$, where the image $\phi(x)$ of an element  of the
fundamental group determines how the sheets are permuted when going around
the closed curve corresponding to $x\in \G_g$. This covering is usually not
connected, its connected components being in one-to-one correspondence
with the orbits $\xi\in\cO(\phi)$ of the permutation group $\phi(\G_g)$ 
acting on $\{1,\dots ,n\}$, where $\cO(\phi)$ denotes the set of these
orbits. The connected component corresponding to the orbit $\xi\in\cO(\phi)$
is itself a closed Riemann surface, whose genus is 
\beq g_\xi=\vert\xi\vert (g-1)+1 \label{RH} \eeq
by the Riemann-Hurwitz formula,
where $\vert\xi\vert$ is the length of the orbit $\xi$. Each such 
connected component gives a contribution $Z_{g_\xi}(\tau_\xi)$ to
the partition function, and all that remains to be done is to determine
the Fricke coordinate $\tau_\xi : \G_{g_\xi}\to SL(2,{\bf R})$ of this
component. To achieve this, select any point $\xi^*\in\xi$ of the orbit
$\xi$, and consider its stabilizer
\beq \cS_\xi=\{ x\in\G_g \ver \phi(x)\xi^*=\xi^*\} \label{stab} \eeq
Note that the stabilizer depends on the actual choice of the representative
$\xi^*$, but the stabilizers corresponding to different representatives of
the same orbit are conjugate subgroups in $\G_g$. It follows from elementary
considerations that $\cS_\xi$ is a subgroup of finite index $\vert\xi\vert$ 
in $\G_g$,
\beq [\G_g : \cS_\xi ]=\vert\xi\vert \label{index} \eeq
But any subgroup of $\G_g$ of index $\vert\xi\vert$ is isomorphic to 
the group $\G_{g_\xi}$ by the general theory of surface groups. This means
that the restriction
\beq \tau_\xi : \cS_\xi\to SL(2,{\bf R}) \label{tau}\eeq
of $\tau$ to the subgroup $\cS_\xi$ may be considered as an embedding
\beq \tau_\xi : \G_{g_\xi}\to SL(2,{\bf R}) \eeq
i.e. it determines a point in the Teichmuller space $\cT_{g_\xi}$, and it
is obvious that this point does not depend on the actual choice of
the orbit representative $\xi^*$. 

Let's note two important features of Eq.\rf{part}. First, to determine
the Fricke coordinate $\tau_\xi$ of the covering surface corresponding to
the orbit $\xi$, we have made use of the existence of an isomorphism
\beq \iota : \cS_\xi\to\G_{g_\xi} \eeq
to identify $\cS_\xi$ with $\G_{g_\xi}$. But, while the theory of surface
groups ensures us of the existence of such an isomorphism, it is by no 
means unique, as the composite map $\iota\circ\alpha$ for arbitrary
$\alpha\in M_{g_\xi}$ is as good as well. This means that it is not a 
point in Teichmuller space, but rather one in the moduli space $\cM_{g_\xi}$
that is determined by the above recipe, so in order for Eq.\rf{part} to make
sense, it is imperative that the partition functions $Z_g(\tau)$ of $\cC$
satisfy the modular invariance requirement Eq.\rf{mod}. The second point is
that the genus $g$ partition function $Z_g^\S$ of $\orbi$ gets contributions
from partition functions of $\cC$ of genus greater than $g$, as a 
consequence of the Riemann-Hurwitz formula Eq.\rf{RH}. The only 
exception to this rule is the case $g=1$,
where we have seen that the genus one partition function of $\orbi$ is 
completely determined by the genus one partition function of $\cC$.

Of course, the significance of Eq.\rf{part} is mostly theoretical, as it is
by no means an easy task to compute explicitly the quantities involved 
- supposing that we know already all the higher genus partition functions 
of $\cC$, which is another matter -, but it should be recognized as an
important step in understanding the relationship between orbifoldization,
covering surfaces and uniformization theory. One may also speculate that
the computation could be performed for some special class of permutation
groups $\S$ - as it is the case for genus one -, the most important 
example being the full symmetric groups $\S=S_n$, where the determination
of the higher genus partition functions of $\orbi$ would yield the higher
loop corrections to the free energy of the second quantized string 
moving in a background described by the CFT $\cC$. 
\bigskip

{\it Acknowledgement :} It is a pleasure to acknowledge discussions with
Zal\'an Horv\'ath, L\'aszl\'o Palla and Christoph  Schweigert.
\bigskip

Work partially supported by OTKA F019477.

\end{document}